# Scattering of inhomogeneous circularly polarized optical field and mechanical manifestation of the internal energy flows

A. Ya. Bekshaev[1*], O. V. Angelsky[2], S. G. Hanson[3], and C. Yu. Zenkova[2]

[1] *I.I. Mechnikov National University, Dvorianska 2, Odessa 65082, Ukraine*

[2] *Chernivtsi National University, Kotsyubinsky Str. 2, Chernivtsi 58012, Ukraine*

[3] *DTU Fotonik, Department of Photonics Engineering, DK-4000 Roskilde, Denmark*

Based on the Mie theory and on the incident beam model via superposition of two plane waves, we analyze numerically the momentum flux of the field scattered by a spherical microparticle placed within the spatially inhomogeneous circularly polarized paraxial light beam. The asymmetry between the forward- and backward-scattered momentum fluxes in the Rayleigh scattering regime appears due to the spin part of the internal energy flow in the incident beam. The transverse ponderomotive forces exerted on dielectric and conducting particles of different sizes are calculated and special features of the mechanical actions produced by the spin and orbital parts of the internal energy flow are recognized. In particular, the transverse orbital flow exerts the transverse force that grows as $a^3$ for conducting and as $a^6$ for dielectric subwavelength particle with radius $a$, in compliance with the dipole mechanism of the field-particle interaction; the force associated with the spin flow behaves as $a^8$ in both cases, which testifies for the non-dipole mechanism. The results can be used for experimental identification and separate investigation of the spin and orbital parts of the internal energy flow in light fields.

## I. INTRODUCTION

The steady interest into light beams with angular momentum and into singular optics [1–4] has stimulated a growing attention to the internal energy flows in light fields (optical currents) [5–12]. The internal energy flow pattern provides a physically meaningful and universal characterization of arbitrary light fields. This is especially suitable for near-field optics and in new applications associated with micro- and nanooptics, invisibility cloaking, superlensing and metamaterials [13–16]. But more important is that the internal energy flows reveal the intimate geometric and dynamic essentials of the light field transformations that underlies any process of a light beam formation, propagation or diffraction [9–12,17]. In particular, the total energy flow density (TFD), represented via the time-average Poynting vector distribution, can be subdivided into the "spin" (SFD) and "orbital" (OFD) parts to reflect the peculiar properties associated with the spin and orbital degrees of freedom of light, especially their unique features and interrelations [9–12]. The spin flow is usually associated with inhomogeneous circular polarization while the orbital one is due to the explicit energy redistribution within an optical beam.

---

[*] E-mail: bekshaev@onu.edu.ua

In view of such important and useful properties, the TFD as well as its spin and orbital constituents appear to be valuable instruments for the light field description and analysis. However, wide utilization of these instruments is hampered by difficulties in their experimental detection (visualization) and measurement. As far as we can judge, the only regular way of energy flow measurement relies upon determining the electric and, if necessary, magnetic vectors of the optical field followed by the Poynting vector calculation via the standard formulas [13,18,19]. In this situation, the possibilities for immediate detection and/or visualization of the internal flows become rather attractive and appealing. The most promising approach, repeatedly used in experimental practice [20–23], is based on the mechanical action the energy flow exerts on probe particles with various sizes, shapes and optical properties. This relies on the fact that the TFD is proportional to the field momentum density [24], and on the assumption that a particle, due to absorption, reflection or scattering of the incident light, partially feels the light field momentum and starts to move in accordance with its local value and direction. During the past years, this approach has been well elaborated, mainly due to close connection with problems of optically driven micro-machines, micro-engineering and micromanipulation [20,25].

Nevertheless, despite many impressive practical results, in applications related to the fundamental study of the internal flows, this method is still far from ideal. The main reason is that the field-induced motion of particles depends on many additional factors. Together with the electromagnetic ponderomotive influences of non-Poynting origin (gradient force, dissipative force, polarization-dependent dipole force [26–29]), the specific ghost effects may occur due to the medium in which the probe particles are suspended (radiometric, photophoretic forces, the medium viscosity, etc.), the particle-containing cell (its configuration, the wall friction) and because of the particle shape and material [20,25]. Even in situations where all non-Poynting sources are isolated (e.g., due to special geometry of the field and the measuring equipment [20,26] or by proper calibration procedures), it is rather difficult to establish an exact numerical correspondence between the probe particle motion and the local value of the field momentum. First, there is no simple and transparent model for a force produced by the electromagnetic momentum interacting with a medium [10], and second, any particle disturbs the electromagnetic field in the ambient region, of at least on the order of a wavelength, and the real perturbed field felt by the particle may be very different from the unperturbed field pattern that existed before the particle was present [12,30,31].

The situation is further complicated by the existence of two sorts of energy flow of different nature. The mechanical action of the OFD can be satisfactorily explained by using the notion of transverse light pressure [12]. At the same time, although the SFD's ability to cause translation or orbital motion of particles has been proved both by simulation [26] and in experiment [32,33], the physical nature of this effect cannot be understood based on the existing model of the SFD origin [3,7]. Moreover, the usual model of the optical force acting on a sub-wavelength sized particle (in the Rayleigh scattering regime), which is based on the classical dipole-interaction Hamiltonian [28,29,34], does not predict any force associated with the SFD. Under these circumstances, the question arises on the nature of the SFD-induced mechanical action and on the mechanical equivalence between the OFD and the SFD: whether the spin and orbital momenta produce the same motion of a particle, provided that they possess the same direction and magnitude?

In our opinion, a possible way of resolving the above issues can be found in considering relatively simple model situations where the relations between the force acting on a particle and the energy flow in the incident optical field can be easily calculated and interpreted, so that contributions of the spin and orbital energy flow constituents can be "isolated" and studied separately. An example of such an approach was described recently [26]: the incident field is formed by only two plane waves, which allows one to employ the standard Mie theory for calculating mechanical action of the field. Subsequently, the results are juxtaposed with the TFD, OFD and SFD patterns of the incident field, as well as with its energy distribution. This enables identification of the ponderomotive contributions owing to the different energy flow constituents together with the influences of the non-Poynting factors with explicit account for the field structure and the particle's optical properties. Despite its simplicity, this model is flexible enough to represent the main features of fields with inhomogeneous distributions of amplitude, phase and polarization that can be realized experimentally [27]. In the present work, within the frame of this model, we "construct" field configurations that distinctly differ by the SFD and OFD patterns and analyze the mechanical actions they exert on dielectric and absorbing particles. This allows us to reveal some special features of the ponderomotive influences associated with the different kinds of field momentum, their similar and distinguishing aspects, and to discuss possibilities for practical detection of the internal energy flows and discrimination between their spin and orbital constituents.

## II. MODEL DESCRIPTION

### A. Incident field

In what follows, we will consider monochromatic optical waves where the electric and magnetic vectors can be written as $\text{Re}[\mathbf{E}\exp(-i\omega t)]$, $\text{Re}[\mathbf{H}\exp(-i\omega t)]$ with complex amplitudes $\mathbf{E}$ and $\mathbf{H}$. Hence the time-average Poynting vector distribution $\mathbf{S}$ representing the field TFD and the field momentum density $\mathbf{p}$ are determined by expression [24]

$$\mathbf{S} = c^2 \mathbf{p} = gc\,\text{Re}(\mathbf{E}^* \times \mathbf{H}). \tag{1}$$

Here $g = (8\pi)^{-1}$ in the Gaussian system of units, $c$ is the velocity of light in vacuum. Due to the proportionality between $\mathbf{S}$ and $\mathbf{p}$ we only operate with one of these quantities in what follows; we choose $\mathbf{p}$, though preserving the name "energy flow density" for its physical meaning. In further consideration we shall take into account that the field-particle interaction takes place in a medium rather than in vacuum; then $\mathbf{p}$ in Eq. (1) represents the kinetic (Abraham) momentum density of the electromagnetic field, and its decomposition into the spin and orbital parts, $\mathbf{p}_S$ and $\mathbf{p}_O$, reads [35]

$$\mathbf{p}_S = \frac{g}{4\omega}\text{Im}\left[\nabla \times \left(\frac{1}{\mu}\mathbf{E}^* \times \mathbf{E}\right) + \nabla \times \left(\frac{1}{\varepsilon}\mathbf{H}^* \times \mathbf{H}\right)\right], \tag{2}$$

$$\mathbf{p}_O = \mathbf{p} - \mathbf{p}_S$$
$$= \frac{g}{2\omega}\text{Im}\left[\frac{1}{\mu}\mathbf{E}^* \cdot (\nabla)\mathbf{E} - \frac{1}{2}\nabla\left(\frac{1}{\mu}\right) \times (\mathbf{E}^* \times \mathbf{E}) + \frac{1}{\varepsilon}\mathbf{H}^* \cdot (\nabla)\mathbf{H} - \frac{1}{2}\nabla\left(\frac{1}{\varepsilon}\right) \times (\mathbf{H}^* \times \mathbf{H})\right] \tag{3}$$

where $\varepsilon$ and $\mu$ are the medium permittivity and permeability, respectively, and $\mathbf{E}^* \cdot (\nabla)\mathbf{E}$ is the invariant notation for the vector operation that in Cartesian components reads [10,12]

$$\mathbf{E}^* \cdot (\nabla)\mathbf{E} = E_x^* \nabla E_x + E_y^* \nabla E_y + E_z^* \nabla E_z.$$

Eqs. (2) and (3) are written for general conditions admitting an inhomogeneous medium; in a homogeneous case, Eq. (3) can be simplified omitting the terms with gradients of $\varepsilon^{-1}$ and $\mu^{-1}$. Another important quantity that plays a substantial role in the analysis is the time-average energy density:

$$w = \frac{g}{2}\left(\varepsilon|\mathbf{E}|^2 + \mu|\mathbf{H}|^2\right). \tag{4}$$

Following Ref. [26], our task is to determine the electromagnetic field perturbed by the presence of a particle, then to calculate its momentum and to compare the result with the initial momentum carried by the unperturbed incident field. In general, due to the particle presence, the scattered field $\mathbf{E}_{sc}$, $\mathbf{H}_{sc}$ emerges that adds to the incident field $\mathbf{E}$, $\mathbf{H}$ [31] causing the total field momentum density to be changed by

$$\Delta\mathbf{p} = \frac{g}{c}\text{Re}\left[(\mathbf{E}^* + \mathbf{E}_{sc}^*) \times (\mathbf{H} + \mathbf{H}_{sc}) - \mathbf{E}^* \times \mathbf{H}\right] = \frac{g}{c}\text{Re}(\mathbf{E}_{sc}^* \times \mathbf{H}_{sc} + \mathbf{E}^* \times \mathbf{H}_{sc} + \mathbf{E}_{sc}^* \times \mathbf{H}). \tag{5}$$

The change of the field momentum results in a recoil force acting on the particle. This force can be determined by deriving the field momentum flux through the spherical surface $A_R$ with radius $R \to \infty$ surrounding the particle. In the medium with the refractive index $n = \sqrt{\varepsilon\mu}$, this force equals

$$\mathbf{F} = -\frac{c}{n}\oint_{A_R} \Delta\mathbf{p}\, dA = -\frac{c}{n} R^2 \oint \Delta\mathbf{p}\, d\Omega \qquad (6)$$

where $d\Omega$ indicates integration over the solid angle. This depends on the particle position, and our aim is to find correspondence between $\mathbf{F}$ and the incident field momentum in the point where the particle is located.

The incident field configuration for our model is illustrated in Fig. 1. The incident light beam comes from the lower hemisphere ($z < 0$) and illuminates a spherical particle whose center is situated at the origin of the laboratory frame ($xyz$). To represent an inhomogeneous field distribution over the nominal transverse plane $z = 0$, the incident field should be formed by a superposition of plane waves differently oriented with respect to the nominal longitudinal axis $z$. The $j$-th plane wave propagates along axis $z_j$ that deviates from the laboratory axis $z$ by the incidence angle $\gamma_j$; in what follows, we restrict ourselves to the case when angles $\gamma_j$ lie in the coordinate plane ($yz$). With each member of the plane-wave superposition, the "proper" coordinate frame ($x\ y_j\ z_j$) is associated, connected with the laboratory frame by the relations

$$z_j = z\cos\gamma_j + y\sin\gamma_j, \quad y_j = -z\sin\gamma_j + y\cos\gamma_j. \qquad (7)$$

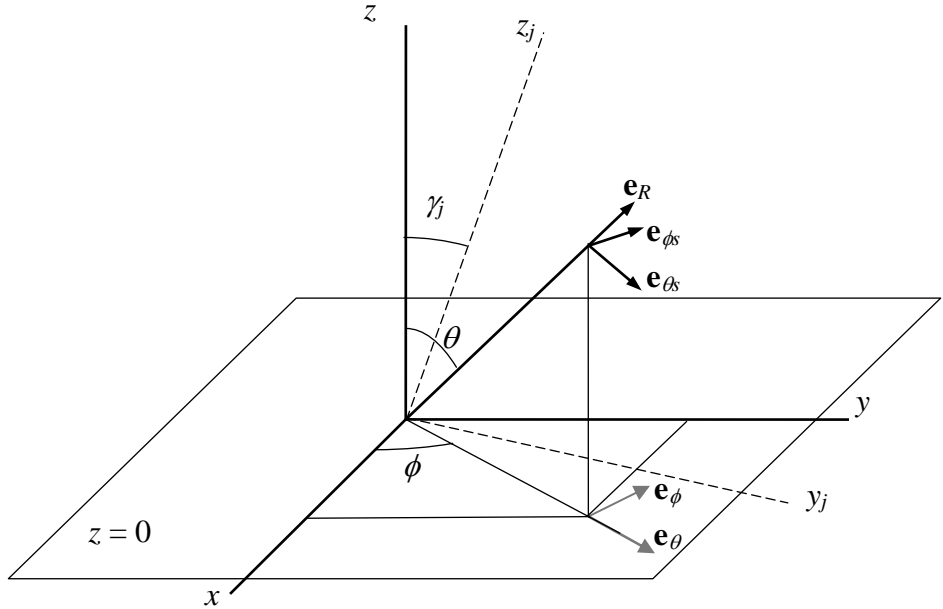

Fig. 1. Geometrical conditions of the light scattering analysis. The particle is situated in the coordinate origin, incident light comes from the lower hemisphere; other parameters are explained in the text.

In its proper frame, the electric and magnetic fields of a separate plane-wave component are described by the equations

$$\mathbf{E}_{aj}(x, y_j, z_j) = \mathbf{E}_j \exp(ikz_j) = \begin{pmatrix} E_{xj} \\ E_{yj} \end{pmatrix} \exp(ikz_j),$$

$$\mathbf{H}_{aj}(x, y_j, z_j) \equiv \mathbf{H}_j \exp(ikz_j) = \sqrt{\frac{\varepsilon}{\mu}}\, \mathbf{e}_j \times \mathbf{E}_{aj}(z_j) \qquad (8)$$

where $E_{xj}$ and $E_{yj}$ are constants, $\mathbf{e}_j$ is the unit vector for the $z_j$-axis and $k = n\omega/c$ is the wavenumber of the incident radiation in the medium. The optical field distribution, created by waves (8) in the common reference plane $z = 0$, is generally inhomogeneous and in the laboratory coordinates can be written in the form

$$\mathbf{E}_{aj}(x, y) = \begin{pmatrix} E_{xj} \\ E_{yj}\cos\gamma_j \\ -E_{yj}\sin\gamma_j \end{pmatrix} \exp(ikz_j), \quad \mathbf{H}_{aj}(x, y) = \sqrt{\frac{\varepsilon}{\mu}} \begin{pmatrix} -E_{yj} \\ E_{xj}\cos\gamma_j \\ -E_{xj}\sin\gamma_j \end{pmatrix} \exp(ikz_j), \quad (9)$$

where $z_j$ is related to $y$ and $z$ by the first Eq. (6).

Here, we restrict ourselves to the case where the superposition consists of only two plane waves ($j = 1, 2$) [36], and the electric and magnetic strengths of the incident optical field are equal to
$$\mathbf{E} = \mathbf{E}_{a1} + \mathbf{E}_{a2}, \quad \mathbf{H} = \mathbf{H}_{a1} + \mathbf{H}_{a2}. \quad (10)$$
Following Ref. [26], one finds the field energy density (4) as well as the SFD (2) and OFD (3) components:

$$w = \varepsilon g \left[ |\mathbf{E}_1|^2 + |\mathbf{E}_2|^2 + \cos^2\frac{\gamma_1 - \gamma_2}{2} D(y, z) \right]; \quad (11)$$

$$p_{Sx} = \frac{g_e}{2c}\sin(\gamma_1 - \gamma_2)\left[\left(E_{x2}^* E_{y1} - E_{y2}^* E_{x1}\right)e^{ik(z_1 - z_2)} + \left(E_{y1}^* E_{x2} - E_{x1}^* E_{y2}\right)e^{ik(z_2 - z_1)}\right], \quad (12)$$

$$p_{Sy} = \frac{g_e}{2c}\sin^2\frac{\gamma_1 - \gamma_2}{2}(\sin\gamma_1 + \sin\gamma_2)D(y, z), \quad (13)$$

$$p_{Sz} = \frac{g_e}{2c}\sin^2\frac{\gamma_1 - \gamma_2}{2}(\cos\gamma_1 + \cos\gamma_2)D(y, z); \quad (14)$$

$$p_{Ox} = 0, \quad (15)$$

$$p_{Oy} = \frac{g_e}{c}\left[|\mathbf{E}_1|^2\sin\gamma_1 + |\mathbf{E}_2|^2\sin\gamma_2 + \frac{1}{2}\cos^2\frac{\gamma_1 - \gamma_2}{2}(\sin\gamma_1 + \sin\gamma_2)D(y, z)\right], \quad (16)$$

$$p_{Oz} = \frac{g_e}{c}\left[|\mathbf{E}_1|^2\cos\gamma_1 + |\mathbf{E}_2|^2\cos\gamma_2 + \frac{1}{2}\cos^2\frac{\gamma_1 - \gamma_2}{2}(\cos\gamma_1 + \cos\gamma_2)D(y, z)\right] \quad (17)$$

where

$$g_e = g\sqrt{\frac{\varepsilon}{\mu}}, \quad D(y, z) = \left(E_{x2}^* E_{x1} + E_{y2}^* E_{y1}\right)e^{ik(z_1 - z_2)} + \left(E_{x1}^* E_{x2} + E_{y1}^* E_{y2}\right)e^{ik(z_2 - z_1)}. \quad (18)$$

Eqs. (11) – (18) show that the simple superposition of Eqs. (10) can serve as a model for a rather general inhomogeneous field with non-zero spin and orbital flows [26]. Note that in the considered field geometry, due to Eqs. (16) and (17), the $x$-component of the OFD is absent and the entire $x$-directed flow is of a spin nature (12).

### B. Scattered field and mechanical action

The light scattered by a spherical particle illuminated by a plane monochromatic wave can be calculated using the Mie theory [31,37]. To find the field mechanical action (5), one should know the scattered field at $R \to \infty$. For such conditions, the scattered field produced by the $j$-th plane wave (9) can be found via the relations

$$\mathbf{E}_{scj} = \frac{e^{ikR}}{-ikR}\mathbf{E}_{sj}, \quad \mathbf{H}_{scj} = \frac{e^{ikR}}{-ikR}\mathbf{H}_{sj} \quad (19)$$

where

$$\mathbf{E}_{sj} = \begin{pmatrix} E_{\theta sj} \\ E_{\phi sj} \end{pmatrix} = \begin{pmatrix} S_2 & 0 \\ 0 & S_1 \end{pmatrix}\begin{pmatrix} E_{\theta j} \\ E_{\phi j} \end{pmatrix} = \begin{pmatrix} S_2 & 0 \\ 0 & S_1 \end{pmatrix}\begin{pmatrix} E_{xj}\cos\phi_j + E_{yj}\sin\phi_j \\ -E_{xj}\sin\phi_j + E_{yj}\cos\phi_j \end{pmatrix},$$

$$\mathbf{H}_{sj} = \begin{pmatrix} H_{\theta sj} \\ H_{\phi sj} \end{pmatrix} = \sqrt{\frac{\varepsilon}{\mu}} \begin{pmatrix} 0 & -1 \\ 1 & 0 \end{pmatrix} \begin{pmatrix} E_{\theta sj} \\ E_{\phi sj} \end{pmatrix}, \qquad (20)$$

$S_1 \equiv S_1(\cos\theta_j)$ and $S_2 \equiv S_2(\cos\theta_j)$ are elements of the scattering matrix [31,37] dependent on the wavenumber $k$, particle radius $a$ and the complex refractive index $m$ of the particle material relative to the ambient medium. In Eqs. (20), the Cartesian and spherical coordinates are measured in the frame $(x, y_j, z_j)$ associated with the $j$-th incident plane wave (cf. Fig. 1). The scattered field is completely transverse, i.e. all the components of Eqs. (20) are orthogonal to the unit vector $\mathbf{e}_R$. In the simplest case of Rayleigh scattering when the particle is much smaller than the wavelength, we have

$$S_1 = -i(ka)^3 \frac{m^2 - 1}{m^2 + 2}, \quad S_2 = S_1 \cos\theta_j. \qquad (21)$$

In more general situations, $S_1$ and $S_2$ are expressed via the spherical vector harmonics [31]. Each plane wave of the incident field is scattered independently, so the resulting scattered field can be found by vector summation of the results obtained from Eqs. (19) and (20). In view of relations (19) and for future convenience, we represent it in the form

$$\mathbf{E}_{sc1} + \mathbf{E}_{sc2} = \mathbf{E}_s \frac{e^{ikR}}{kR}, \quad \mathbf{H}_{sc1} + \mathbf{H}_{sc2} = \mathbf{H}_s \frac{e^{ikR}}{kR}. \qquad (22)$$

The auxiliary quantities $\mathbf{E}_s$ and $\mathbf{H}_s$ represent the "meaningful" parts of the scattered field amplitudes – slowly varying envelopes imposed over the standard spherical wave factor $e^{ikR}/kR$.

Now, with allowance for Eqs. (10), Eq. (5) can be written as

$$\Delta\mathbf{p} = \frac{g}{ckR}\mathrm{Re}\left[\frac{\mathbf{E}_s^* \times \mathbf{H}_s}{kR} + \mathbf{E}_1^* \times \mathbf{H}_s e^{ik(R-z_1)} + \mathbf{E}_2^* \times \mathbf{H}_s e^{ik(R-z_2)} + \mathbf{E}_s^* \times \mathbf{H}_1 e^{-ik(R-z_1)} + \mathbf{E}_s^* \times \mathbf{H}_2 e^{-ik(R-z_2)}\right], \qquad (23)$$

which should be substituted into Eq. (6). The first term in the brackets describes the momentum of the scattered field and attracts separate interest. For further references, it would be suitable to represent its contribution to the ponderomotive force in Eq. (6) as a sum of two terms that express the momentum flux into the forward $\mathbf{P}(+)$ and backward $\mathbf{P}(-)$ hemisphere, respectively:

$$\mathbf{F}_s = -\frac{g}{n(kR)^2} \oint_{A_R} \mathbf{E}_s^* \times \mathbf{H}_s dA = -\mathbf{P}_s(+) - \mathbf{P}_s(-) \qquad (24)$$

where

$$\mathbf{P}_s(\pm) = \frac{g}{\mu k^2} \int_0^{2\pi} d\phi \int_{\pi/4(1\mp 1)}^{\pi/4(3\mp 1)} \left(|E_{\theta s}|^2 + |E_{\phi s}|^2\right) \mathbf{e}_R(\theta, \phi) \sin\theta d\theta; \qquad (25)$$

the latter expression is derived using Eqs. (20).

In principle, the "scattering-field force" of Eq. (24) can be evaluated analytically, at least in the Rayleigh regime when Eqs. (21) are fulfilled; however, the presence of two plane waves (10) and the necessity to switch between the three coordinate frames makes this way too cumbersome. Fortunately, due to the well developed methods for calculation of the scattering matrix (20) [31], the problem can easily be solved numerically. This is not the case for the other terms of Eq. (23) whose numerical integration is practically impossible because of the oscillating interference factors $\exp[\pm ik(R - z_j)]$. Nevertheless, their expressions for $R \to \infty$ can be found analytically via the asymptotic approximation of the integral

$$\int_0^{\pi} U(\theta) e^{\pm ikR(1-\cos\theta)} \sin\theta d\theta = \frac{e^{\pm ikR}}{\mp ikR}\left[e^{\pm ikR}U(\theta)_{\theta=\pi} - e^{\mp ikR}U(\theta)_{\theta=0}\right] + O\left(\frac{1}{k^2 R^2}\right)$$

where $U(\theta)$ is an arbitrary function with sufficiently regular behavior. When applied to summands of Eq. (23), this formula gives

$$R^2 \text{Re} \int_0^{2\pi} d\phi_j \int_0^{\pi} \left[ \mathbf{E}_j^* \times \mathbf{H}_s e^{ik(R-z_j)} + \mathbf{E}_s^* \times \mathbf{H}_j e^{-ik(R-z_j)} \right] \sin\theta_j d\theta_j$$

$$= R^2 \frac{2\pi}{kR} \text{Im}\left[ e^{2ikR} \left(\mathbf{E}_j^* \times \mathbf{H}_s\right)_{\theta_j=\pi} - \left(\mathbf{E}_j^* \times \mathbf{H}_s\right)_{\theta_j=0} - e^{-2ikR}\left(\mathbf{E}_s^* \times \mathbf{H}_j\right)_{\theta_j=\pi} + \left(\mathbf{E}_s^* \times \mathbf{H}_j\right)_{\theta_j=0} \right]. \quad (26)$$

Further, since $\mathbf{H}_j = \mathbf{e}_{zj} \times \mathbf{E}_j$ and $\mathbf{H}_s = \mathbf{e}_R \times \mathbf{E}_s$, the following relations fulfill:

$$\mathbf{E}_s^* \times \mathbf{H}_j = \mathbf{E}_s^* \times \left(\mathbf{e}_{zj} \times \mathbf{E}_j\right) = \mathbf{e}_{zj}\left(\mathbf{E}_s^* \cdot \mathbf{E}_j\right) - \mathbf{E}_j\left(\mathbf{e}_{zj} \cdot \mathbf{E}_s^*\right),$$

$$\mathbf{E}_j^* \times \mathbf{H}_s = \mathbf{E}_j^* \times \left(\mathbf{e}_R \times \mathbf{E}_s\right) = \mathbf{e}_R\left(\mathbf{E}_j^* \cdot \mathbf{E}_s\right) - \mathbf{E}_s\left(\mathbf{e}_R \cdot \mathbf{E}_j^*\right). \quad (27)$$

Note that at position $\theta_j = 0$ $\mathbf{e}_R = \mathbf{e}_{zj}$, and at $\theta_j = \pi$ $\mathbf{e}_R = -\mathbf{e}_{zj}$ (cf. Fig. 1). Accordingly, in these points the second terms of Eqs. (27) vanish, and

$$\left(\mathbf{E}_j^* \times \mathbf{H}_s\right)_{\theta_j=0} = \left(\mathbf{E}_s^* \times \mathbf{H}_j\right)_{\theta_j=0}^*, \quad \left(\mathbf{E}_s^* \times \mathbf{H}_j\right)_{\theta_j=\pi} = -\left(\mathbf{E}_j^* \times \mathbf{H}_s\right)_{\theta_j=\pi}^*. \quad (28)$$

As a result, the contributions at points $\theta_j = \pi$ in expression (26) tend to zero and, combining Eqs. (26), (28), (23) and (6), one obtains

$$\mathbf{F} = \mathbf{F}_s - g \frac{4\pi}{nk^2} \text{Im}\left[ \left(\mathbf{E}_s^* \times \mathbf{H}_1\right)_{\theta_1=0} + \left(\mathbf{E}_s^* \times \mathbf{H}_2\right)_{\theta_2=0} \right] \quad (29)$$

where $\mathbf{F}_s$ is given by Eqs. (24) and (25), $\mathbf{E}_s$ is determined by Eqs. (22), (19) and (20), and $\mathbf{H}_j$ is the amplitude of the incident plane-wave component as defined in Eq. (8). Note that due to the accepted incident field geometry (Fig. 1), both terms in bracket in Eq. (29) are vectors belonging to plane $(yz)$, and the $x$-component of the total force exerted on the particle, $F_x = F_{xs}$, is fully determined by the scattered-field force (24).

### III. MECHANICAL PROPERTIES OF THE SCATTERED FIELD

The model described by Eqs. (9) and (11) – (18) is applicable to a number of practical situations involving incident fields with inhomogeneous amplitude-phase profile and polarization [27]. For the detailed analysis, we choose the simplest configurations that enable the analysis of the special features of the SFD and TFD [26]. The first example is realized by a symmetric superposition of circularly polarized plane waves (9) that appears if

$$\gamma_1 = -\gamma_2 = \gamma, \quad (30)$$

$$\begin{pmatrix} E_{x1} \\ E_{y1} \end{pmatrix} = E_0 \begin{pmatrix} 1 \\ \sigma i \end{pmatrix} \quad (31)$$

and both waves are identical with possible phase shift, i.e.

$$\begin{pmatrix} E_{x2} \\ E_{y2} \end{pmatrix} = \exp(i\delta) \begin{pmatrix} E_{x1} \\ E_{y1} \end{pmatrix} \quad (32)$$

where $\sigma = \pm 1$ is the polarization helicity (spin number). In the paraxial case ($\gamma \ll 1$), to which we are restricted in this paper, terms $\sim \gamma^2$ and of higher order can be neglected, and Eqs. (11) – (18) reduce to [26]

$$w = 4\varepsilon g |E_0|^2 (1 + \cos 2\Phi), \quad (33)$$

$$p_{Sx} = -\frac{4g_e}{c} \sigma\gamma |E_0|^2 \sin 2\Phi, \quad (34)$$

$$p_{Sy} = p_{Sz} = p_{Oy} = p_{Ox} = 0,$$

$$p_{Oz} = \frac{4g_e}{c} |E_0|^2 (1 + \cos 2\Phi) \quad (35)$$

with

$$\Phi = \gamma ky - \frac{\delta}{2}. \tag{36}$$

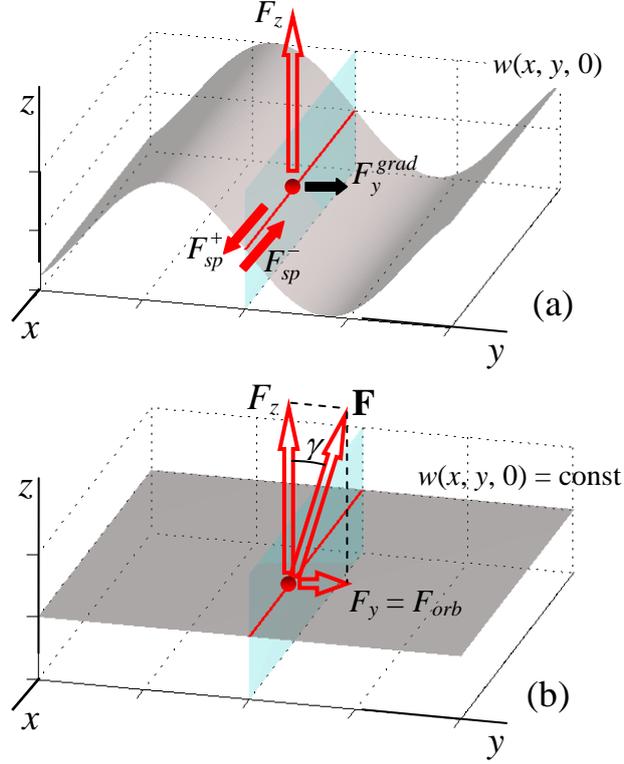

Fig. 2. Mechanical forces acting on a probe particle situated in plane $z = 0$ and illuminated by the superposition of two plane waves with (a) different directions, resulting in the inhomogeneous circularly polarized field described by Eqs. (30) – (35) and (b) coinciding directions, resulting in a spatially homogeneous energy distribution but having inclined wavefront, according to Eqs. (44), (45). Further explanations are found in the text.

These equations characterize the field with inhomogeneous energy distribution over the reference plane $z = 0$ [Fig. 2(a)]. The OFD (35) of this field is directed along the propagation axis $z$ and is completely longitudinal; the internal OFD is absent. As should be expected under the paraxial conditions [12], the longitudinal momentum (35) and the energy density (33) are connected by the standard relation $p_z = nw/c$. The longitudinal momentum produces the usual light pressure force $F_z$ pushing the particle forward; the inhomogeneous intensity is the source of the gradient force $F_y^{grad}$ [(see Fig. 2(a)], which was analyzed elsewhere [26]. Both these forces are not the subjects of the present consideration; instead, we intend to concentrate on the transverse SFD (34), which represents the most interesting feature of the discussed field model. The very appearance of the $x$-directed transverse momentum in this field geometry seems counter-intuitive, though it immediately follows from the spin flow theory [7] and is quite expectedly [9,12] oriented along the constant-energy lines $y = $ const.

Its physical nature is partially elucidated by considering the scattering field momentum flux behavior with increasing particle size parameter

$$\xi = ka, \tag{37}$$

where $a$ is the particle radius. We calculated the scattered field parameters of Eqs. (24) and (25) for two sorts of spherical particles suspended in water ($\varepsilon = 1.77$, $\mu = 1$, $n = 1.33$): metallic (gold in water, relative refraction index $m = 0.32 + 2.65i$ [37]), and dielectric (latex in water, $m = 1.12$); the radiation wavenumber is $k = 1.33 \cdot 10^5 \text{cm}^{-1}$ (He-Ne laser). The results obtained for the field model of Eqs. (30) – (32) are given by the curves $P_{sx}^+(\pm)$ in Fig. 3; note that only results for the field model with $\sigma = +1$ in Eq. (31) are presented because switching the polarization helicity to $\sigma = -1$ causes nothing but the sign reversal, $P_{sx}^-(\pm) = -P_{sx}^+(\pm)$. For comparison, in Fig. 3 the results are also presented obtained for different cases of the linearly polarized incident beam, which occur if, instead of Eq. (31), the following relations take place:

$$E_{y1} = 0, \; E_{x1} = E_0 \neq 0 \tag{38}$$

("x-polarization", curves $P_{sx}^x(\pm)$ and $P_{sy}^x(\pm)$),

$$E_{x1} = 0, \; E_{y1} = E_0 \neq 0 \tag{39}$$

("y-polarization", curves $P_{sx}^y(\pm)$ and $P_{sy}^y(\pm)$), and

$$E_{y1} = E_{x1} = E_0 \neq 0 \tag{40}$$

("45° polarization", curves $P_{sx}^{45}(\pm)$ and $P_{sy}^{45}(\pm)$). Upon calculations, condition

$$2\Phi = -\pi/2 \tag{41}$$

($y = 0$, $\delta = \pi/2$ or $\delta = 0$, $y = \pi/(4k\gamma)$) was chosen which corresponds to the maximum absolute value of the spin flow (34); the angle between the two interfering plane waves was assumed to equal

$$\gamma = 0.01 \text{ rad.} \tag{42}$$

To eliminate the influence of the incident beam intensity and to decrease the range of presented data, these are normalized by dividing the calculated quantities by the total momentum flux of the incident field through the particle cross section, which in the considered field configurations of Eqs. (30) – (32) and (38) – (40) equals

$$P_0 = \frac{2g}{\mu}\left(|E_{x1}|^2 + |E_{y1}|^2\right)(1 + \cos 2\Phi) \cdot \pi a^2. \tag{43}$$

Additionally, each curve in Fig. 3 represents the fourth root of the corresponding momentum flux value.

The curves $P_{sx,sy}^{\cdots}(+)$ and $P_{sx,sy}^{\cdots}(-)$ in Fig. 3 represent the transverse Cartesian components of the front and rear half-sphere momentum fluxes calculated via Eq. (25) for differently polarized fields. Usually, in the small-particle limit (Rayleigh scattering regime [31]), the scattering is considered symmetric, in agreement with Eqs. (21). This implies that the scattered radiation carries no transverse momentum flux; and in fact, for linearly polarized incident light, the results of Fig. 3 confirm this suggestion with rather high accuracy, at least until $\xi = 0.2$, where the particle size influence becomes perceptible. Even beyond this range limit, the forward and backward scattered momentum fluxes are almost similar for all cases with linearly polarized incident field as presented in Fig. 3.

However, when the particle is illuminated by circularly polarized light, both the forward- and backward-scattered momentum fluxes possess noticeable x-directed components [curves $P_{sx}^+(\pm)$ in Figs. 3(a), (b)], much more intense than any analogous contribution emerging under the linearly-polarized illumination. According to Fig. 3 and with allowance for the normalization factor (43), quantities $\left|P_{sx}^+(\pm)\right|$ grow as $a^6$ with the particle size, which is of the same order as is the total scattered intensity [31], and which strongly exceeds the scattered field anisotropy that may appear in a linearly polarized field. Besides, the forward $P_{sx}^+(+)$ and backward $P_{sx}^+(-)$ transverse momentum fluxes essentially differ, even by the sign. These effects can be attributed to the optical vortex generation (spin-to-orbital angular momentum conversion) upon scattering of light with circular polarization [38,39], which directly follows from Eqs. (20) and (21). In the laboratory frame, the forward-scattered and backward-scattered vortices possess opposite helicities, and this is the source of the scattering asymmetry. For a single incident plane wave, this asymmetry is "hidden" in the hemisphere momentum fluxes due to integration over the azimuth angle $\phi$ in Eq. (25) but interference of the scattered fields produced by the two plane waves makes it visible. One can expect the revealed forward-backward asymmetry to be a general feature inherent in scattering of any spatially inhomogeneous elliptically polarized beams in the Rayleigh scattering regime [26]. This effect can be considered as a novel manifestation of the spin-orbit interaction upon light scattering, whose observation requires a spatially inhomogeneous incident field.

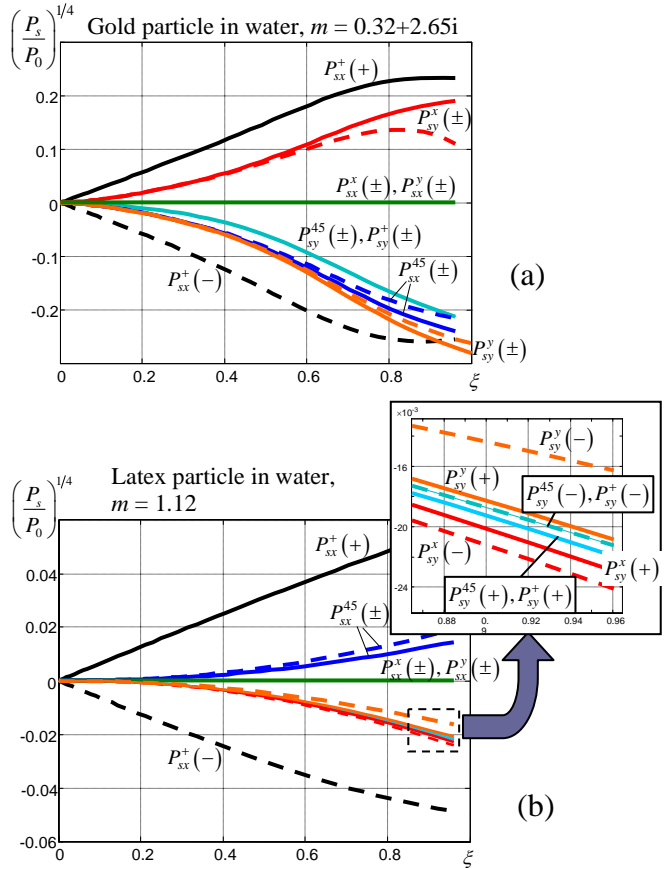

Fig. 3. Fourth root of the normalized momentum flux $\mathbf{P}_s/P_0$ [see Eqs. (25) and (43)] components of the field scattered by a spherical (a) metallic and (b) dielectric particle, suspended in water, vs the particle size parameter (37), calculated for the conditions of Eqs. (41), (42). Each curve is marked by the corresponding component notation: subscripts ($x$, $y$) denote the momentum flux Cartesian component, superscripts ($x$, $y$, 45, +) denote the incident field polarization as indicated by Eqs. (31) and (38) – (40). Solid (dashed) lines describe the momentum flux into the forward (backward) hemisphere; the inset shows magnification of the dashed rectangle in panel (b). In both cases (a) and (b), curves $P_{sx}^x(\pm)$ and $P_{sx}^y(\pm)$ coincide with the zero line, curves $P_{sy}^{45}(\pm)$ and $P_{sy}^{+}(\pm)$ visually merge; however, the small difference between the forward and backward scattered contributions can be traced in the inset.

What is even more important, the absolute values of $P_{sx}^+(+)$ and $P_{sx}^+(-)$ are not exactly the same, and their difference is just the source of the "net" scattered field momentum flux that can be determined via Eq. (24). This momentum flux is "balanced" by the recoil force exerted on the particle – the physical reason for the translational ponderomotive influence of the inhomogeneous circularly polarized optical field, which can be treated as the mechanical action of the spin energy flow [26]. The comparative study of this mechanical action is the subject of the next Section.

## IV. COMPARISON OF THE MECHANICAL ACTIONS OF THE SPIN AND ORBITAL ENERGY FLOWS

In this section, we apply the general procedure formulated in Section IIB, Eqs. (24), (25) and (29), to the calculation of the mechanical forces exerted on the probe particles and confront it with the energy flow pattern in the incident circularly polarized field. Recently, this procedure was described in detail [26], and now we follow it with no essential modifications.

An impressive feature of the calculated results is the "counter-intuitive" *x*-directed force represented by curves $F_{sp}^{\pm}$ in Fig. 4(a), (b) and by arrows $F_{sp}^{\pm}$ in Fig. 2(a): apparently, the field configuration of Eqs. (8) – (10) and (30) – (32) looks symmetric with respect to the *x*-axis reversal, and it is the "invisible" instant field vector rotation that destroys this symmetry. Moreover, this force changes sign upon switching the polarization handedness and vanishes in linearly polarized fields[1] [which directly follows from near-zero values of the partial momentum fluxes $P_{sx}^x(\pm)$, $P_{sx}^y(\pm)$ presented in Fig. 3(a), (b)]. All this is akin to the behavior of the SFD (34) which, in accordance with Eqs. (33) – (35), represents the only *x*-directed energy flow contribution emerging in the incident field specified by Eqs. (30) – (32); besides, there is no energy gradient in the *x*-direction. That is why it is quite natural to associate the *x*-directed force $F_{sp}^{\pm}$ with the SFD of the incident field and consider it as the mechanical action of the spin energy flow.

The next step is to compare this force with the orbital flow action. Unfortunately, in the field that satisfies Eqs. (30) – (32), the transverse OFD is absent, and the really existing *y*-directed force [Fig. 2(a)] was identified as the gradient force [26], which is useless for our purpose of analyzing the mechanical action owing to the internal energy flows. To avoid this inconvenience, one could address more complex field configurations in which the orbital and spin internal flows are present simultaneously (see, e.g., Refs. [9,10]). However, in such situations not only the force calculation will be much more cumbersome and time-consuming, but also the interpretation of the numerical results and separation between the SFD-induced and OFD-induced force contributions becomes rather difficult.

A more promising way is to find a simple field configuration in which the OFD-induced action $F_{orb}$ can be easily identified, and its correspondence with the orbital flow distribution can be established. If, additionally, the local OFD value is numerically equal to the SFD in the model of Eqs. (31) – (36), then the calculated forces $F_{orb}$ and $F_{sp}^{\pm}$ can be treated as "pure" manifestations of the orbital and spin flow mechanical actions, respectively. The fact that $F_{orb}$ and $F_{sp}^{\pm}$ are realized in different optical fields is not important because, if there exist any regularities of the ponderomotive effects inherent in the SFD (OFD) *per se*, their manifestations should be identical provided that the local SFD (OFD) values are the same, regardless of all other field parameters and details.

Formally, a simple configuration with obviously identifiable OFD action can be realized based on the same two-plane-wave field of Eqs. (10) – (18). All one must do is to accept, in contrast to Eq. (30), the condition

---

[1] In fact, when the linear polarization of the incident field differs from "pure" *x*- or *y*-polarization, the specific *x*-directed force appears but it is not related to any energy flow constituent and can be attributed to the polarization inhomogeneity [26].

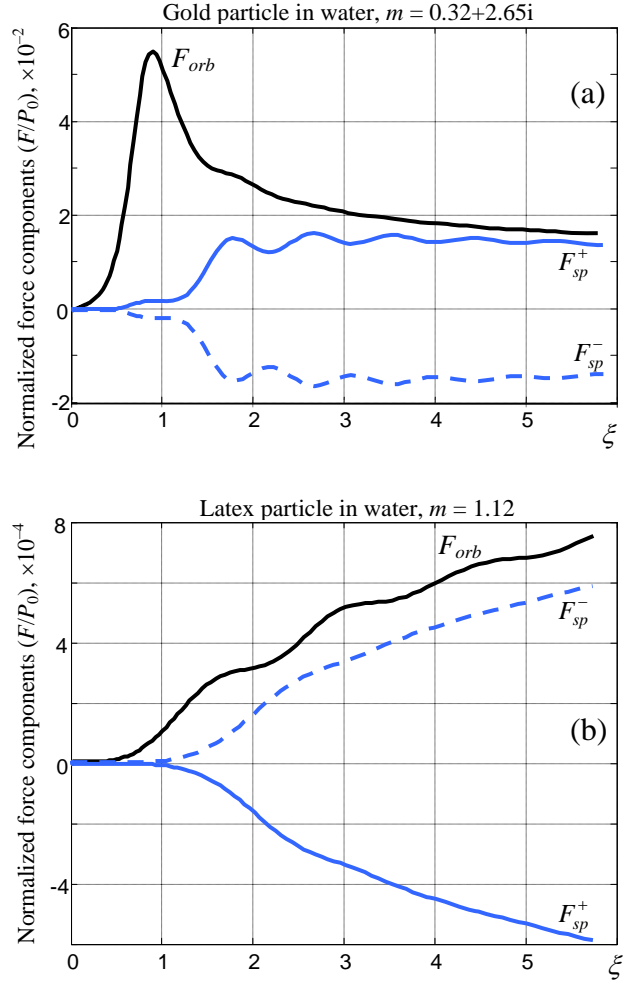

Fig. 4. Comparison of the mechanical actions associated with the spin and orbital internal energy flows for (a) metal and (b) dielectric spherical particle suspended in water. Each curve represents the force dependence of the particle size parameter (37). The SFD-induced forces $F_{sp}^{\pm}$ are calculated as *x*-components of the force experienced by a particle in the field characterized by Eqs. (30) – (36) for the conditions of Eqs. (41), (42) [cf. Fig. 2(a)]; their signs change upon switching the polarization handedness from $\sigma = 1$ (solid lines) to $\sigma = -1$ (dashed lines). The OFD-induced contributions $F_{orb}$ are determined as the *y*-components of the force exerted on a particle in the field characterized by Eqs. (44) and (41), (42) [cf. Fig. 2(b)].

$$\gamma_1 = \gamma_2 = \gamma. \tag{44}$$

Here the SFD vanishes and the whole transverse momentum is of orbital nature and equals

$$p_{Oy} = \frac{4g_e}{c}\gamma|E_0|^2(1+\cos\delta), \tag{45}$$

which signifies that, keeping the same values of $\gamma$, $E_0$ and choosing $2\Phi = -\delta$ [according to Eq. (36), this is realized, e.g., if $y = 0$], the *y*-directed OFD (45) in the field configuration of Eqs. (44), (31) and (32) equals the *x*-directed SFD (34) in the field configuration of Eqs. (30) – (32). Hence, one can verify the equivalence and/or discrepancies between the mechanical actions of the spin and orbital energy flows merely by juxtaposing the *x*-directed force $F_{sp}^{\pm}$ calculated as described above, and the *y*-directed force $F_{orb}$ calculated for the incident field configuration specified by Eqs. (44),

(31) and (32). Note that in this situation the incident momentum flux through the particle cross section,

$$P_0 = \frac{2g}{\mu}\left(|E_{x1}|^2 + |E_{y1}|^2\right)(1 + \cos\delta) \cdot \pi a^2 = \frac{4g}{\mu}|E_0|^2 (1 + \cos\delta) \cdot \pi a^2, \quad (46)$$

numerically coincides with the normalizing divider (43). Thus the normalization of $F_{orb}$ via division by the quantity (46) is compatible with the normalization of $F_{sp}^{\pm}$ by (43) and the normalized force values presented in Fig. 4 provide a correct comparison.

One may notice that condition (44) reduces the two-plane-wave superposition to the case of a single plane wave that approaches the nominal observation plane $z = 0$ at a small angle $\gamma$. In this field, the energy is distributed homogeneously and the OFD (45) is, in fact, the transverse projection of the plane-wave momentum directed normally to the wavefront [Fig. 2(b)]. This circumstance does not limit the generality of the consideration: it complies with the common notion that the OFD represents the transverse energy transportation of the same nature as the "main" longitudinal energy flow [12], and that the OFD-induced mechanical action is nothing but a sort of transverse light pressure. Such origination of the transverse ponderomotive action is reflected in Fig. 2(b) by decomposition of the light-pressure vector **F** into the longitudinal $F_z$ and transverse $F_y = F_{orb}$ components.

In view of the above remarks, the specific features of the SFD- and OFD-induced mechanical actions can be investigated via the comparative analysis of curves $F_{sp}^{\pm}$ and $F_{orb}$ in Figs. 4(a), (b). The first and quite expected difference between the two is that, both for metallic [Fig. 4(a)] and dielectric [Fig. 4(b)] particles, the spin-flow contribution changes sign together with the polarization handedness (curves $F_{sp}^{+}$ and $F_{sp}^{-}$), while the orbital flow action does not. Noteworthy, in case of dielectric particles the spin-induced force is directed oppositely to the spin energy flow (the similar behavior was predicted recently for dielectric particles in air [26]). For example, Eq. (34) dictates that $p_{Sx}$ is positive at $\sigma = 1$ and under accepted conditions $y = 0$, $\delta = \pi/2$, whereas curve $F_{sp}^{+}$ of Fig. 4(b) lies in the negative half-plane. This feature is associated with the specific spatial asymmetry of the scattered momentum, which can be seen by comparison of curves $P_{sx}^{x}(+)$ and $P_{sx}^{x}(-)$ in Fig. 3. For dielectric particles, Fig. 3(b), the absolute contribution of the front half-sphere [curve $P_{sx}^{x}(+)$] is a bit higher than that of the rear half-sphere [curve $P_{sx}^{x}(-)$] whereas for conducting particles, Fig. 3(a), the reverse relation is observed.

In all other aspects, the similarities and discrepancies of the spin- and orbital-flow actions are not immediately recognized. Both contributions depend on the particle size and optical properties, and conditions are possible when one of the discussed contributions is prevailing: for example, very small particles ($\xi \ll 1$) "feel" the OFD much stronger than the SFD is felt. Probably, the difference in the $F_{sp}^{\pm}$- and $F_{orb}$- dependences on $\xi$ can be used for separate detection of the spin and orbital energy flows.

The curves in Fig. 4 demonstrate the general rule that any light field's mechanical action is essentially mediated by the probe particle size and properties. In this context, the results for very small particles are more representative since their dependence on the particle size and refractive index is more regular and is not modified by resonance phenomena [31]. The corresponding data are presented in Fig. 5. Among other things, they reveal the difference between the OFD action exerted on the dielectric ($F_{orb} \sim a^6$) and the metallic ($F_{orb} \sim a^3$) particles. This can be explained by the prevailing role of light absorption in the light-pressure effect on conductive particles, whereas the OFD-induced action experienced by a dielectric particle is completely due to elastic scattering [31]; the same reason (in combination with the small relative refractive index) underlies the appreciably lower absolute force values in case of dielectric particles, cf. the vertical scales in Figs.

4a and 4b. Another important observation involves the peculiar regularity of the SFD-induced action, which for both considered cases satisfies the relation $F_{sp}^{\pm} \sim a^8$. Undoubtedly, this feature is related to the special mechanism of the SFD-induced ponderomotive effect and reflects its particular physical nature.

Interestingly, in the dipole approximation [28,29,34], the entire mechanical action experienced by Rayleigh particles consists of the gradient force (first summand of Eq. (5) of Ref. [34]) and the OFD-induced force (the "scattering force" – second summand of Eq. (5) of [34] – is proportional to the "electric" part of the OFD, without restoring the "electric-magnetic democracy" [10]). In fact, the apparently spin-dependent "field gradient force" {Eqs. (6) and (8) of [34]}, named also "curl force associated to the nonuniform distribution of the spin angular momentum" {Eq. (11) of Ref. [29]} can be introduced but it merely cancels out the contribution of $\mathbf{p}_S$ [(Eq. (2)] contained in the TFD-proportional term which Simpson and Hanna call "dissipative radiation force" (Eq. (7), Ref. [34]) and Albaladejo et al. term as the "traditional radiation pressure" (Eq. (13), Ref. [29]). As a result, the genuine SFD-induced mechanical action escapes from the known dipole-based calculations. In this context, the particle-size dependence of the OFD-induced forces seen in Fig. 5 ($\sim a^3$ for conductive and $\sim a^6$ for dielectric particles) is completely justified by the behavior of the imaginary part of the polarizability for absorbing and non-absorbing particles (Eq. (11) of [34]).

The eighth power dependency on the particle size testifies that the SFD-induced mechanical action is not of dipole nature and only appears in higher degrees of the multipole expansion. Note however that the relative weakness of the spin-flow force in the Rayleigh-scattering region does not mean that it is always smaller than the one associated with orbital flow: as Fig. 4 testifies, at $\xi > 2$, both forces are quite comparable.

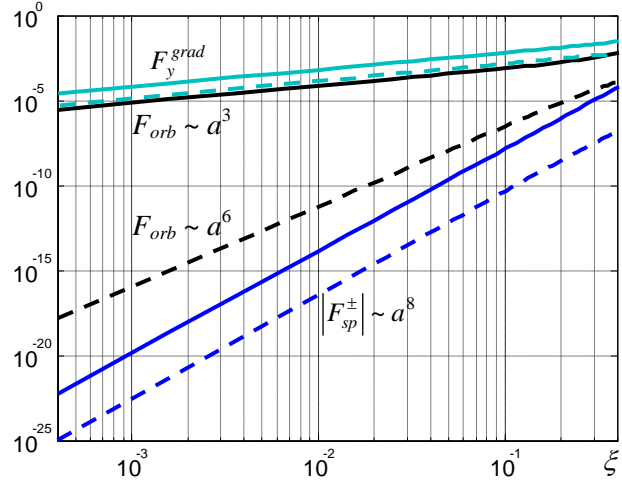

Fig. 5. Initial segments of curves, presented by Fig. 4, in double logarithmic scale. Solid lines: metallic particle [Fig. 4(a)], dashed lines: dielectric particle [Fig. 4(b)]. Orders of the force growth with the particle radius $a$ are indicated with allowance for the normalization factor $P_0$ (43). For comparison, the behavior of the gradient force $F_y^{grad}$ [Fig. 2(a)] is included.

In view of these facts, recent suggestions on the mechanical equivalency of the spin and orbital energy flows [26] should be essentially corrected, if not rejected. Really, both the spin and orbital contributions to the energy flow are able to cause translational motion of the probe particles but the quantitative characters of the spin-induced and the orbital-induced motions and their dependences on the particle size and properties are rather different. Accordingly, the SFD and OFD can experimentally be distinguished employing probe particles with deliberately chosen sizes and properties. Such a choice requires a detailed analysis of the expected behavior for various sorts of particles, which can be performed based on the approach presented in this paper.

## V. CONCLUSION

The results of this work are based on a model of a spatially inhomogeneous optical field [26] that is formed by superposition of two plane waves. Despite its simplicity, the model adequately represents some general properties of inhomogeneous fields, including the main regularities inherent in internal energy flow and its spin and orbital parts. By using the Mie theory, mechanical characteristics of the field scattered by a probe particle placed within the spatially inhomogeneous circularly polarized light beam are studied. In particular, the forward – backward scattering field asymmetry, observable in the Rayleigh scattering regime (for particle sizes much less than the radiation wavelength) and associated with the spin energy flow of the incident beam, has been revealed and analyzed. This spin-flow induced asymmetry is closely related to the known effect of the spin-orbital angular momentum conversion for Rayleigh scattering [38,39], and can be treated as a novel manifestation of the spin-orbit interaction of light.

By means of numerical calculations, the ponderomotive forces exerted on spherical microparticles with conductive and dielectric properties, exposed to light fields with different configurations, are also investigated. Two specific incident field configurations were considered in detail: (i) spatially inhomogeneous beam with the well-defined transverse spin flow and (ii) spatially homogeneous field with a transverse orbital flow (inclined plane wave). Comparison of the ponderomotive actions, performed in both cases, permitted us to disclose the special features of the mechanical action inherent in the spin and orbital parts of the internal energy flow. In particular, for the sub-wavelength (Rayleigh) particles, the orbital-flow force grows as $a^3$ for conducting and as $a^6$ for dielectric particle with radius $a$, in compliance with the dipole interaction mechanism [28,34]; the spin-flow force appears in higher multipole orders and behaves as $a^8$ in both cases. As well, our simulations show that, for any particle sizes, the spin flow may "pull" dielectric particles against its own direction whereas the orbital flow always "pushes" probe particles along the energy transportation lines. These differences reflect the unique ways in which the spin and orbital momenta of light are transmitted to material bodies, in particular, the essential role of non-dipole interactions in the spin flow mechanical action.

We hope that the results of this work can be useful for experimental identification and separate investigation of the spin and orbital parts of the internal energy flow in light fields via the probe particle's motion. Concurrently, our results indicate some difficulties and limitations of the approaches based on the ponderomotive action and probe bodies. Primarily, there are a number of "parasite" sources of mechanical influences that are not related with the field momentum or any of its constituents and usually mask their contributions. Among various factors mentioned in the Introduction, here we emphasize the non-Poynting source of the electromagnetic origin that plays an important role in any inhomogeneous light field: the gradient force. According to Fig. 5 (the lines $F_y^{grad}$) the gradient force, generally, exceeds the energy-flow-induced contributions, and this circumstance should be properly addressed in any experiment (see also Refs. [26,28,29]). Another essential issue is that beyond the Rayleigh-scattering range of the probe particle sizes, the electromagnetic field-induced mechanical action depends on the particle radius and on the complex refractive index in a rather complex and apparently irregular fashion, strongly affected by the field inhomogeneity, and this makes it very difficult to establish an accurate numerical correspondence between the transverse ponderomotive force and the internal energy flow component which, theoretically, gives rise to this force. In most cases, the resulting force value is rather far from the naïve expectations that the mechanical action is proportional to the local value of the incident field momentum density in the point where the particle is placed [26]. In essence, in such situations not the field momentum (energy flow) *per se* but the electromagnetic field as a whole acts as a motive power, and it seems doubtful that the observable force experienced by a particle can be definitely associated with the transverse field momentum at all.

Finally, we emphasize that the model of an inhomogeneous optical field described in this paper, even in its simplest version, provides consistent deductions related to the mechanical actions of spatially inhomogeneous vector light fields. The presented model can easily be generalized to describe more complicated situations to reflect additional fine features of the real optical fields.